# A pathway towards proper modeling of physical properties

K. Koperwas[1, 2,*], A. Grzybowski[1, 2], and M. Paluch[1, 2]

[1] Institute of Physics, University of Silesia in Katowice, 75 Pułku Piechoty 1, 41-500 Chorzów, Poland

[2] Silesian Center for Education and Interdisciplinary Research, SMCEBI, 75 Pułku Piechoty 1a, 41-500 Chorzów, Poland

*corresponding author: kajetan.koperwas@smcebi.edu.pl

**ABSTRACT:**

Theoretical concepts in condensed matter physics are typically verified and also developed by exploiting computer simulations mostly in simple models. Predictions based on these usually isotropic models are often at odds with measurement results obtained for real materials. On the other hand, all-atom simulations are complex and time-consuming. In this paper, we formulate a new strategy for effective molecular modelling, which properly reflects properties of real particles by using quasi-real molecules of simple but anisotropic architecture and identifying the applicable range of intermolecular interactions for a given physical process or quantity. As a demonstration of our method capabilities, we solve an intriguing problem within the density scaling idea that has attracted attention in recent decades due to its hallmarks of universality. It demonstrates that the new strategy for molecular modelling opens broad perspectives for simulation and theoretical research, for example, into unifying concepts in the glass transition physics.

**MANUSRIPT:**

For over half a century till the present time a tremendous scientific effort has been made to understand the nature of liquids. Undeniably, some breakthroughs in the research on the liquid-state were from computer simulations of modelled systems to explore the behaviours of real materials.[1–3] Most of the computational experiments were devoted to the so-called simple-liquids (i.e., systems composed of many particles interacting via radially symmetric pair potential), exhibiting some properties similar to real liquids. However, the simplicity of the molecular architecture of simple-liquids as well as the intermolecular interactions assumed raises the question of how well these model systems can capture the intricate thermodynamic and dynamics properties of real liquids.

Among numerous computational studies of simple-liquids since the latter part of the past century, prevalent are those performed with the intermolecular interaction described by the Lennard-Jones (LJ) potential (and its approximation by the soft-sphere potentials valid for short distances between molecules). The reason for the unceasing interest in the LJ systems is the availability of theoretical explanation of the origins of the repulsive and attractive parts of the potential. The repulsive interaction is caused by the overlapping of the electron clouds at short distances, whereas the attractive interaction results from electrostatic interactions between permanent as well as induced dipole moments.[4] Hence, the use of the LJ potential (or its approximations) enables rational modelling of the physical interactions between molecules of real van der Waals liquids. Consequently, throughout the last decades, many processes occurring in systems described by LJ or soft-sphere potential have been systematically examined. The research revealed the fascinating quasi-universality of the properties[5–10] among which is the thermodynamic scaling of properties based on nature of the potential.[11–14]

For the soft-spheres with the inverse power law (IPL) potential, a rigorous result derived from theory is that all structural, dynamic, and transport properties at equilibrium for a given system are functions of the product variable[15-19], $T(V)^{m/3}$, where $T$ is the temperature, $V$ is the volume of the system, and $m$ is the power of the IPL potential. Since $m$ characterizes the intermolecular potential, the $T(V)^{m/3}$-scaling makes the connection between micro and macro properties of the system. Additionally, the form of $T(V)^{m/3}$-scaling implies that the microscopic information about intermolecular interactions can be obtained from the analysis of macroscopic properties, which is remarkably informative in the context of real materials for which on the intermolecular potential or its parameter is unknown. At this point it should be mentioned that despite $T(V)^{m/3}$-scaling is valid only for soft-sphere systems, it was shown that it is a reasonable approximation also for LJ systems at low temperatures, and hence at high densities[20-24], and thus even for real van der Waals liquids. Therefore, the first experimental observation of quasi-universality in real liquids firmly maintains the fascination of researchers on this subject, and make it one of the most frequently studied issues of condensed matter physics over the last twenty years.[25-29] Initial experimental works on the quasi-universality of real liquids were reported by Tölle[30,31], who analysed quasielastic neutron scattering data of ortho-terphenyl (OTP), the canonical van der Walls liquid, and pointed out that the observed dynamic crossover could be characterized by an effective constant value of $Tv^4$, where $v$ denotes the specific volume. The power $4$ was immediately related via $m/3$ to the exponent 12 of the LJ repulsive core. Motivated by the result of Tölle, Dreyfus successfully scaled rotational relaxation times, obtained from light-scattering data for different isotherms of OTP, onto a single master curve as a function of $Tv^4$.[32] These very promising results have initiated intensive studies of the molecular dynamics in the universality of the thermodynamic-scaling (or the density-scaling) for various real materials. However, it had been quickly pointed out by a few research teams that the power of 4 in the density-scaling found for OTP is not the same for many tested glass formers.[33–36] Consequently, more general scaling rule were formulated[33–40], according to which a dynamic quantity, $X$ (e.g., structural relaxation time $\tau$, viscosity, $\eta$, or diffusion constant, $D$) can be expressed as a function $f$,

$$X = f(Tv^\gamma), \tag{1}$$

where the density-scaling exponent, $\gamma$, is a material constant. At the same time, parallel theoretical studies made impact on the understanding of the origin of observed differences between $\gamma$ and its expected value of 4 from the LJ potential. According to these studies, $\gamma$ is not directly related to the repulsive part of LJ potential, but to its effective approximation at small distances by the IPL potential, $U_{IPL}(r) = \varepsilon(\sigma/r)^m + A$, where $A$ is a constant. The reason why exponent $m$ can differ from 12 is because of a presence of the attractive part of the potential (different for various materials), which causes that the slope $m$ of the effective potential at small distances to differ from that of its pure repulsive core.[22–24,28,41–43] Consequently, it explains why the scaling exponent $\gamma = m/3$ is material dependent,

The theoretical explanation of the density-scaling together with the experimental observation of it for more than 100 liquids and polymers[44] have made scaling of the molecular dynamics from the normal liquid state to near glass transition an inherent part of the supercooled liquid physics. However, the complete understanding of its nature is still at large because of the following problem. The soft-sphere potential, which lies at the heart of the density-scaling, also implies mutual dependence of the thermodynamic quantities. Based on the IPL potential, Bardik et al. derived an equation for a dependence of pressure on volume at isothermal conditions,[45]

$$p^{conf} - p_0^{conf} = B\left[\left(\frac{V_0}{V}\right)^{\gamma_{EOS}} - 1\right], \qquad (2)$$

where $p^{conf} = p - Nk_B T/V$ is a configurational pressure, $k_B$ the Boltzmann constant, $N$ the number of molecules in a system, $\left(p_0^{conf}, V_0\right)$ the configurational pressure and the volume of a chosen reference state, $B$ a temperature dependent parameter, and $\gamma_{EOS} = m/3$. It is worth mentioning that the above isothermal form of the equation of state (EOS) has been successfully tested for model systems[42] as well as for real substances.[46] Hence, the result of $\gamma_{EOS} = m/3$ from the EOS relationship, implies that $\gamma_{EOS} = \gamma$ for systems obeying equation (2) and the density-scaling, irrespective of either model or real liquids. However, it was subsequently reported that the equality between $\gamma_{EOS}$ and $\gamma$ is not fulfilled for real materials.[46–48] On the other hand, $\gamma_{EOS} = \gamma$ was only observed for model systems.[42] These facts lead to three very fundamental questions. (i) Why the model systems do not exhibit the inconsistency between $\gamma$ and $\gamma_{EOS}$, and consequently to which extent the model systems can mimic the behaviour of real liquids? (ii) Why the independent analysis of dynamic and thermodynamic data lead to different values of the exponent of the soft-sphere potential describing the same intermolecular interactions? (iii) Which parameter, $\gamma$ or $\gamma_{EOS}$, is associated properly with $m$? Hence, the disclosure of the reason for the disagreement between $\gamma_{EOS}$ and $\gamma$ in real liquids is a key to understand not only the origin of density scaling, but also the limitation of applicability of simple-liquids. The latter is especially essential because it impacts on the most essential mission of computer simulations, which is the microscopic understanding of the physics of real materials by capturing their properties observed experimentally.

Providing answers to all questions posed above is the main aim of this paper. Based on the new strategy for molecular modelling, we find the solution to the long-standing problem of the inconsistency between $\gamma_{EOS}$ and $\gamma$, as well as determining which one is directly related to the intermolecular potential. Hence, our work critically contributes to the better understanding of the physics of real substances by showing to which extent the real liquids can be treated as simple-liquids. As a consequence, the simulation method can be considered to be groundbreaking because it indicates the new direction of future computational studies by designing model systems to provide the missing link between the studies on simple-liquids and complex all-atom simulations.

The most significant advantage of simple-liquids is the simplicity of their intermolecular interactions, making them convenient for theoretical study and, some of their features can be directly validated by the intermolecular potential assumed as mentioned before. However, the use of the spherically symmetric potential implies that molecules are treated as spheres, which might cause some critical properties of real molecules not captured or mimicked. In particular, the anisotropy of the molecular shape, and consequently the anisotropy of the molecular interactions, could play an important role in thermodynamic and dynamic properties of van der Waals liquids. Thus possibly the experimentally observed differences between $\gamma_{EOS}$ and $\gamma$ is related to molecular shape anisotropy. This hypothesis can be verified using computer simulations of molecular dynamics of systems specially designed. Molecules used in simulations should exhibit the shape anisotropy, but at the same time they should be as simple as possible, which ensures that the influence of other factors is minimized. Meeting the conditions set forth, we created molecules comprised of four identical atoms arranged in a rhombus shape, see Figs. 1a and 1b. Additionally, to increase the anisotropy of intermolecular interactions of one system, we added opposite charges to atoms placed along the longer diagonal of the molecule, which results in the creation of a dipole moment, $\mu$. Hence we obtained two systems of rhombus-like molecules (RLM), which to some extent duplicate the flat and asymmetric shape of the real molecules and the presence of dipole moment.

In Figs. 1a and 1b the dependence of configurational pressure on molar volume, $v_{mol}$, are presented for the two systems. The values of $\gamma_{EOS}$ determined directly from fitting simulation data to equation (2) are 8.96 and 9.58 respectively for RLM molecules with and without dipole moment. It is worth noting that equation (2) was generalized to the form describing temperature-pressure dependence of volume[49], in which the physical meaning of parameter $B$ has been explained, i.e., $B(T) = B_T^{conf}\left(p_0^{conf}\right)/\gamma_{EOS}$, where $B_T^{conf}$ is the configurational isothermal bulk modulus, $B_T^{conf} = B_T - NkT/V$. Since $\gamma_{EOS}$ is connected with the isothermal bulk modulus $B_T$, this parameter is related to the compression of the system. Moreover, equation (2) implies some scaling property of volumetric data, i.e., in isothermal conditions, $\log\left(\left(p^{conf} - p_0^{conf}\right)/B + 1\right)$ is a linear function on $\log\left(v_0/v_{mol}\right)$ with a slope equal to $\gamma_{EOS}$. As presented in Figs. 1c and 1d, this scaling of volumetric data is very well satisfied for both RLM systems.

The values of the density-scaling exponent, $\gamma$, can be determined from the analysis of diffusivity. Equation (1) implies that the density-scaling exponent is the slope of the linear dependence of $\log(T)$ on $\log(v)$ at constant $D$ presented in the insets of Fig. 2. As we show in Fig. 2, the estimated values of $\gamma$ (which are equal to 6.10 for RLM without dipole moment, $\mu = 0$, and 5.03 for those possessing $\mu \neq 0$) enable an almost perfect density-scaling of $D$ with $\log(1/D)$ fall onto one master curve independently of thermodynamic conditions. However, most interesting is the fact that $\gamma_{EOS} > \gamma$ for RLM. Hence, in contrast to the simple-liquids, RLM behaves like real van der Waals liquids. Moreover, by additionally increasing the anisotropy of interactions, the presence of $\mu$ makes the difference between $\gamma_{EOS}$ and $\gamma$ even larger. Therefore, we can infer that the anisotropy of the molecular shape, and thus the anisotropy of their intermolecular interactions, is the cause of the disagreement between $\gamma_{EOS}$ and $\gamma$.

Notwithstanding, at this point we would like to draw attention to perhaps a puzzling result of our studies, which is the fact that the system with dipole moment $\mu$ has smaller $\gamma$ than the system without the. This fact seems to be counter-intuitive because dipole-dipole interactions are attractive, and thus the presence of $\mu$ makes the slope of the repulsive part of the LJ potential steeper.[50] Consequently, one can expect an increase in $\gamma$ with the presence of $\mu$. However, this reasoning is applicable only for simple-liquids because it does not take into account anisotropy of intermolecular interactions. To fully consider the discussed effect, the intermolecular potential dependence on a distance to the molecule as well as on its orientation need to be determined. It must be stressed that in fact many-body interactions are considered (for the two RLM we get 16 different interactions), which drastically complicates the problem. Therefore, we make the following simplification. Since $\gamma$ is related to the diffusion of the centre of mass of the molecule, the intermolecular potential governing this process can be expressed as a function of the distance between the centre of the given molecule and a particular point in the space, $r_{CM}$. For simplicity, we reduce the choice of the points in the space, to only three most characteristic directions relative to RLM, i.e., along both diagonals of RLM and perpendicular to the plane determined by them. Then, the effective potential between two centres of RLM, $U(r_{CM})$, is the arithmetic average of the three considered potentials.

In Fig. 3a it can be seen that the shape of $U(r_{CM})$, is in fact less steep for molecules with $\mu$. Hence, the lowering of $\gamma$ with an increase in the attractive intermolecular interactions by the gain in $\mu$ can be explained using the proposed simplification of many-body interactions potential. Our analysis reveals also that the addition of dipole-dipole interactions makes the intermolecular potential along the longer diagonal of RLM becoming indeed steeper in the studied density range (result not presented). However, at the same time, the potentials along the other two directions are on 'the attractive part', which overcome the aforementioned effect of interactions along the longer diagonal of RLM, and result in a flatter shape of the mean potential. Consequently, the anisotropy of intermolecular interactions is a reason for the observed decrease in $\gamma$ with the increase in the intermolecular attraction.

Since the behaviour of $U(r_{CM})$ in the presence of $\mu$ is consistent with the manner of $\gamma$ and not $\gamma_{EOS}$ (we recall that $\gamma_{EOS}$ increases with $\mu$), the use of IPL potential with $m = 3\gamma$ to approximate $U(r_{CM})$ is more suitable than that with $m = 3\gamma_{EOS}$. As one can see in Fig. 3a, the employment of $m = 3\gamma$ indeed results in better (and almost perfect) descriptions of the effective interactions potentials. The values of $m$ equal to $3\gamma_{EOS}$ lead to shapes of IPL potentials, which are too steep and thus not accurately describing $U(r_{CM})$. Hence, the relationship $m = 3\gamma$ is true, with the caveat that the effective intermolecular potential must be taken into account for complex molecules. The real two-body intermolecular interactions potential cannot be identified with the interactions potential for real van der Waals liquids. Nevertheless, the consideration of molecular architecture could lead to useful approximations.

The relationship between $\gamma$ and effective intermolecular potential explains the theoretical basis of the density scaling for real molecules. In this context, the confirmation of the relation, $\gamma_{EOS} = m/3$, presents a challenge, although EOS (derived on the assumption of this relationship) works perfectly as shown in Fig. 1. However, one could note that the way to estimate the resultant potential, which was suggested in the previous paragraph, could be not always valid, especially for the main interactions governing the system volume. The hint to the above conjecture can be given by the already mentioned connection between $\gamma_{EOS}$ and the bulk modulus, the physical quantity characterizing the resistance of the system to the compression. In contrast to the diffusion, during compression, not only mutual positions of molecules change but also the distances between them. When the one molecule moves towards another, the repulsive forces exerted on it become stronger, and the most gain in repulsion is observed for atom closest to the other molecule (at least for potentials with power law for the repulsive term). Hence, not only the distances between molecules but also their orientations (positions of the nearest atoms) are essential for the interactions determining the changes of volume. The values of the resultant intermolecular potential between the centres of one molecule and the average position of the nearest atom of another molecule, $U(r_{NA})$, for both types of RLM are shown in Fig. 3b. We can observe that contrary to $U(r_{CM})$ the steepness of $U(r_{NA})$ increases with the addition of $\mu$, which interestingly mirrors the trend of $\gamma_{EOS}$ not $\gamma$. However, the most important conclusion obtained from Fig. 3b is the fact that the approximations of $U(r_{NA})$ using the IPL potential with $m = 3\gamma_{EOS}$ almost perfectly follow the calculated values of the effective potential, while the values of $\gamma$ are not large enough to reproduce the steepness of $U(r_{NA})$ as $m = 3\gamma$. Thus, the relationship $\gamma_{EOS} = m/3$ turns out to be true as well. The difference between $\gamma$ and $\gamma_{EOS}$, which has been pointed out for real materials, originates from the fact that the effective potential (which is the approximation of the many-body interactions) is different for different physical processes, and therefore both examined relationships, $m = 3\gamma$ and $m = 3\gamma_{EOS}$, are valid.

The new approach to molecular dynamics simulations proposed by us has enabled to solve the crucial problem of the quasi-universality of density scaling for real liquids. Presented results suggest that the anisotropy of intermolecular interactions, which invariably occurs for real molecules, is the reason for the discrepancy between the scaling exponents $\gamma \neq \gamma_{EOS}$. Interestingly, the many-body interactions potential can be simplified to the two-body interactions potential. However, the key to this operation is to consider not only molecular architecture but also the physics of processes under investigation. Only then the parameter characterizing a given process (e.g., $\gamma$ or $\gamma_{EOS}$) could be directly related to the effective intermolecular potential. The latter implies different $m$ values of its approximation, and then $\gamma \neq \gamma_{EOS}$. However, the explanation of the reason for $\gamma \neq \gamma_{EOS}$ is not possible without applying the new strategy for molecular modelling and simulations. The point is to project quasi-real molecules, which exclusively exhibit the examined property of real particles, keeping at the same time the modesty of molecular architecture and interactions as much as possible. The roles played in the physics of liquids by the molecular shape, the orientations of their dipole moments, or the stiffness of their bonds, are only a few examples for possible studies, which are not accessible in the case of simple-liquids. Therefore, our pioneer approach provides a missing link between simulations of simple-liquids and real molecules. Consequently, this promising alternative approach for current computer simulation studies of the real liquids opens broad perspectives for further computational research, which could bring many benefits.

**FIGURES:**

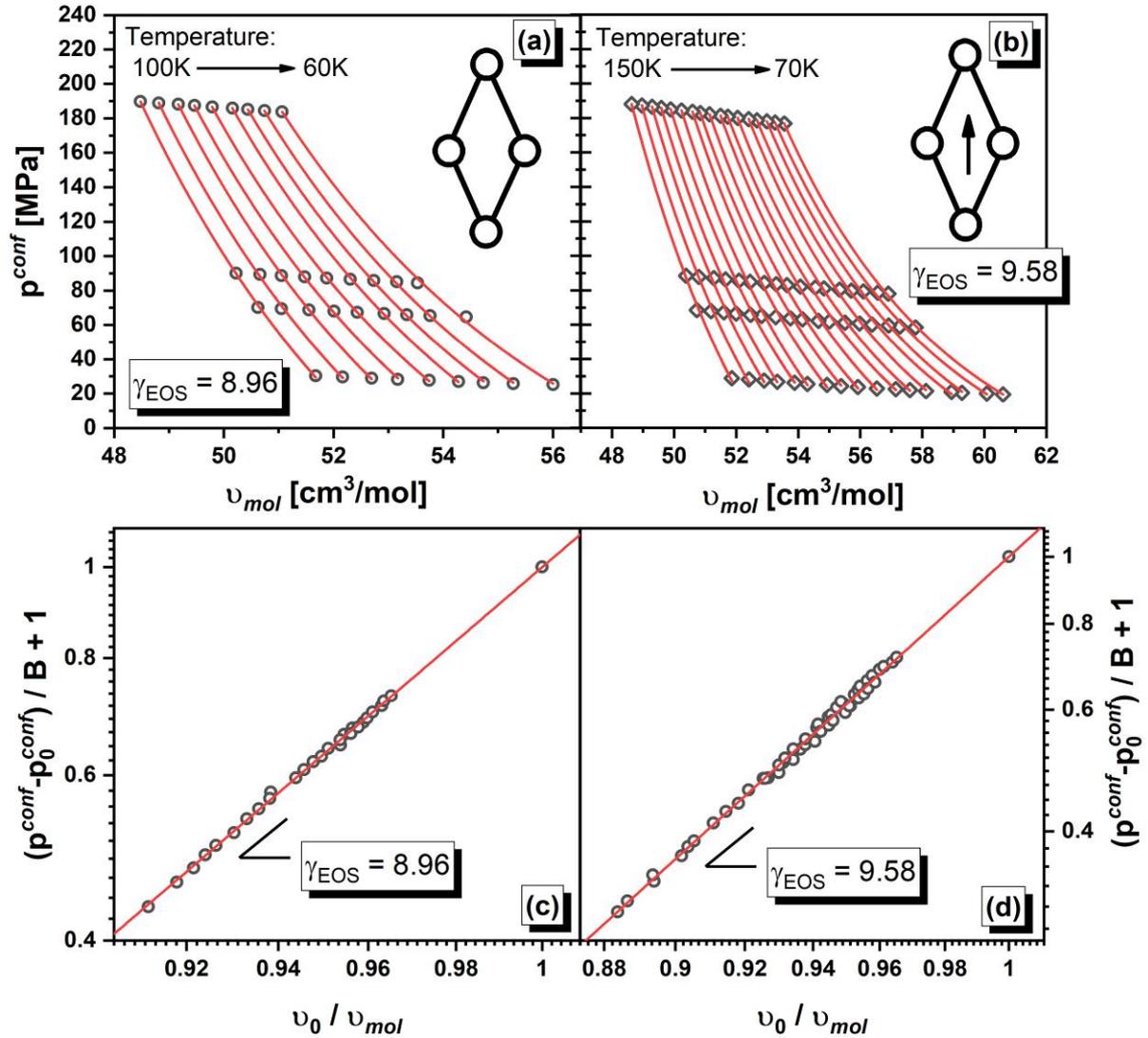

Fig. 1. The isothermal dependence of configurational pressure on molar volume is presented in Figs. 1a and 1b respectively for RLM which does not posses dipole moment and those with the dipole moment. The scheme of the RLM construction is shown in upper right corner of Figs. 1a and 1b (arrow depict dipole moment). The scaling of volumetric data resulting from equation (2) is show in Figs. 1c and 1d for both type of RLM. The slope of the linear dependence is identified with $\gamma_{EOS}$ parameter.

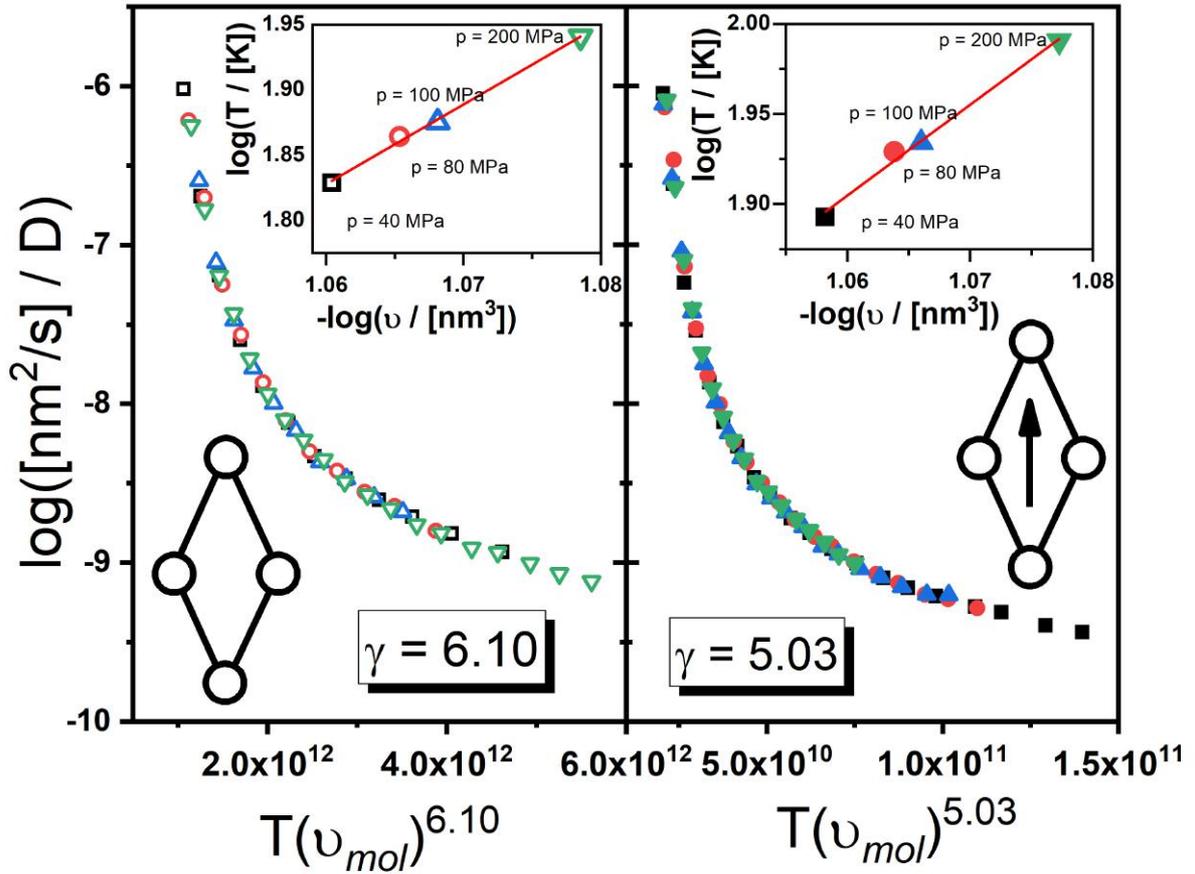

Fig. 2. The density scaling of diffusion coefficients is presented for RLM without dipole moment in Fig 2a and for RLM with dipole moment in Fig 2b. The values of scaling constant, $\gamma$, are obtained from the log-log dependence of temperature on volume at conditions of constant diffusion ($\log\left([\text{nm}^2/\text{s}]/D\right) = -8$), which are presented in insets of Figs. 2a and 2b. The red line represents linear fit of $\log T\left(-\log \upsilon_{mol}\right)$, the slope of which is equal to density scaling constant, $\gamma$.

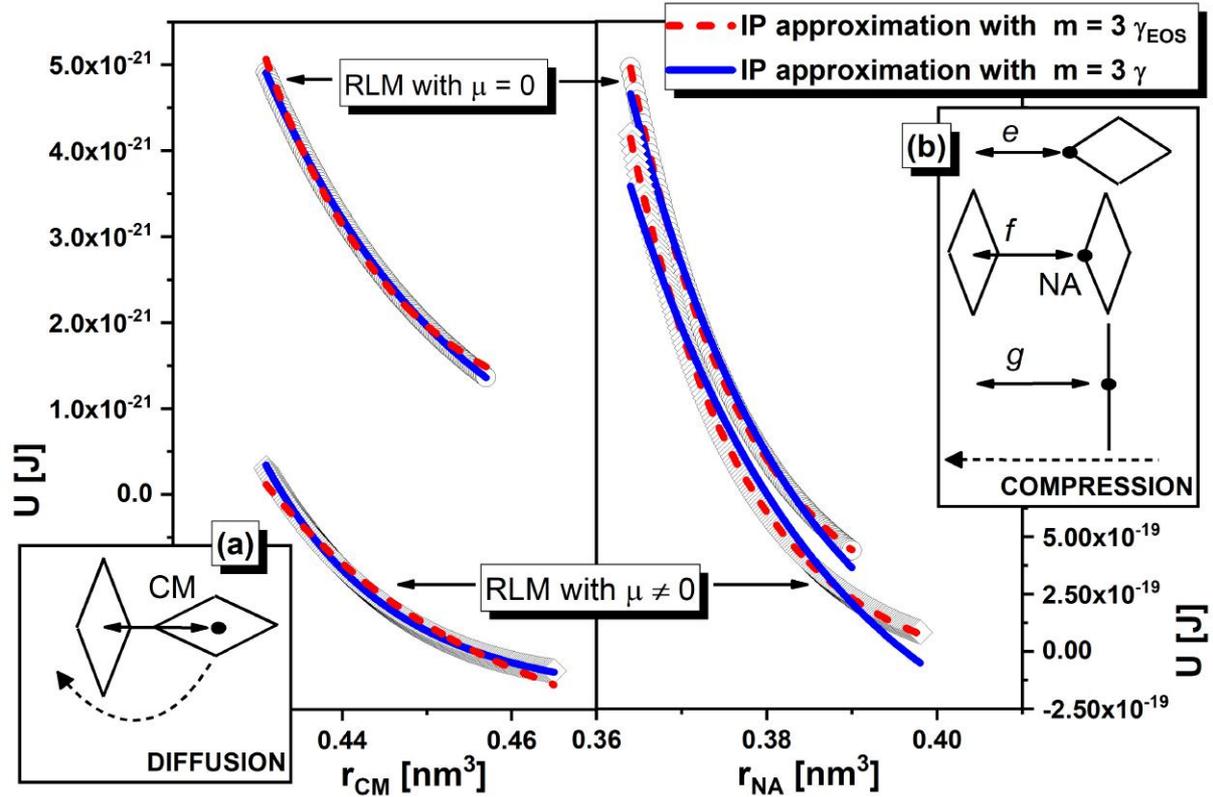

Fig. 3. The resultant interactions potentials for main interactions governing the diffusion process (Fig. 3a) and changes of the volume during compression (Fig. 3b) are presented. $r_{CM}$ denotes the distance between two centers of adjacent molecules, whereas $r_{NA}$ is a distance between center of one RLM and the nearest atom of other RLM. Solid and dashed lines represent fit of the inverse power potential with different values of $m$. Schemes of $U(r_{CM})$ and $U(r_{NA})$ calculation for one orientation of molecule (, which potential is analyzed) are presented in frames. $e$, $f$, $g$ are the distances between the center of one molecule and the nearest atom of neighboring particle

**METHODS:**

In Fig. 1a and 1b the scheme of RLM architecture are presented. RLM are built of four identical atoms, which possess carbon atom mass. The bonds lengths are identical and equal to 0.14982nm (0.14nm is a bonds length for carbon atoms in benzene ring), whereas angles between bonds are set to make one diagonal two times longer than the other. It ensures rhombus shape of molecules and consequently the anisotropy of their structure and interactions in three dimensions. The stiffness of bonds, angles and dihedrals, as well as the non-bonded interaction between atoms of different molecules, have been set by OPLS all-atom force field[51] parameters provided for carbon atoms of the benzene ring. To keep simplicity of constructed quasi-real molecules as much as possible, we decided not to add hydrogen atoms. That, in turn implied the redefinition of atom charges, which was set to $0.0e$ ($e$ is an elementary charge) for RLM without dipole moment. $\mu$ has been introduced by the change of the charge of appropriate atoms to $\pm 0.25e$.

We have used the GROMACS software[52], to perform standard simulations of the molecular dynamics for systems composed of 2048 RLM. Each simulation run was conducted at constant temperature and pressure controlled respectively by the Nose-Hoover thermostat and Martyna-Tuckerman-Tobias-Klein barostat and last for a relatively long time, i.e, 1 bilion of time-steps ($dt = 0.001ps$). The first half of simulation was devoted for equilibration of the system, whereas the volumetric and dynamic data have been collected for the last 500 millions time-steps. Determinations of $\gamma_{EOS}$ and $\gamma$ values began from the construction of perfect FCC lattice crystal, in which the RLM were inserted in the place of atoms. Subsequently, we heated the systems at constant pressure (equal to 40, 80, 100, and 200 MPa) from the starting temperature equals 10K up to the temperature 50K higher than the temperature at which step increase in the volume was observed, indicating the melting of the crystal structure. Then we started the main stage of our experiment, i.e., we performed isobaric cooling ended in the starting temperature. It is worth noting that during this process we observed the characteristic for the glass transition by changes in the slopes of the temperature dependences of the volume, which take place at temperatures for which logarithm of inverse diffusion constant of molecules, $\log\left(1/D \cdot \left[nm^2/s\right]\right)$, are slightly higher than -6. The values of diffusion constant ware calculated using GROMACS software. To avoid analysis of not well-equilibrated or non-equilibrium systems, for further studies we consider only those thermodynamic conditions, which are characterized by $\log\left(1/D \cdot \left[nm^2/s\right]\right) < -6$. Subsequently, the obtained temperature dependences of $\log\left(\left[nm^2/s\right]/D\right)$ have been approximated by the well-known Vogel-Fulcher-Tammann equation, which enable precise estimation of temperatures for a given value of diffusion.

This simplification used for estimation of $U(r_{CM})$ implies a calculation of three different potentials (respectively for particular directions of RLM). Each potential results from the summation of the LJ interactions originating from the atoms of RLM. As it was mention in the main text, we assumed that the orientation of the molecule on which forces are exerted is not important for the diffusion process, which is related to the motion of the centers of the molecules. Hence, the point in space in which the potential is determined could be identified with center molecule, see inset of Fig. 3a. Then the mean potential, $U(r_{CM})$, is the arithmetic average of three considered potentials for RLM (note that in Fig. 3a only one case of shorter diagonal of RLM is presented). For molecules which possess $\mu$ we enhanced all potentials with dipole-dipole interactions according to the well-known formula, $U(r) = -2\mu^4 / 3(4\pi\varepsilon_0)^2 r_{CM}^6 kT$, where average T equal to 110K is used and $\varepsilon_0$ denotes the dielectric permittivity of the vacuum).[4]

The procedure used to estimate $U(r_{NA})$ is more complicated because mutual orientations of both interacting molecules must be taken into account. Considering the three most characteristic orientations of interacting molecules, we can estimate the average distance between center of the one molecule and the nearest atom of the other molecule, $r_{NA}$. The schematic picture is presented in the inset of Fig. 3b. It can be observed that for estimation of $r_{NA}$ the orientation of the molecule, which exerts forces on approaching particle, is not necessary. Hence, we can consider only three cases, which leads to $r_{NA} = (e + f + g)/3$, where $e$, $f$, $g$ are the distances between the center of molecule and the nearest atom of neighboring particle, see Fig. 3b. Nevertheless, the orientations of both molecules must be taken into account for estimation of the effective potential. Then $U(r_{NA})$ can be calculated as an arithmetic average of nine different potentials resulted from three characteristic orientations of each molecule (the dipole-dipole interactions could be calculated according to the formula presented in the previous paragraph). It should be also mentioned that in Fig. 3 the range of $r_{CM}$ (which implies also range of $r_{NA}$) is limited to the reasonable values, which results from the mean volume of single molecules at studied thermodynamic conditions, $r_{CM} = (V/N)^{1/3}$.

**AUTHOR CONTRIBUTIONS:**

K.K. performed the molecular dynamics simulations and data analysis as well as wrote the main manuscript text. A. G. and M. P. contributed to the writing of the manuscript. All authors discussed the results.

**REFERENCES:**

1. Rosenbluth, M. N. & Rosenbluth, A. W. Further results on Monte Carlo equations of state. *J. Chem. Phys.* **22,** 881–884 (1954).

2. Wood, W. W. & Jacobson, J. D. Preliminary results from a recalculation of the Monte Carlo equation of state of hard spheres. *The Journal of Chemical Physics* **27,** 1207–1208 (1957).

3. Alder, B. J. & Wainwright, T. E. Phase Transition for a Hard Sphere System. *J. Chem. Phys.* **27,** 1208–1209 (1957).

4. Atkins, P. & de Paula, J. *Atkins' Physical Chemistry*. (Oxford University Press, 2014).

5. Rosenfeld, Y. Relation between the transport coefficients and the internal entropy of simple systems. *Phys. Rev. A* **15,** 2545–2549 (1977).

6. de J. Guevara-Rodriguez, F. & Medina-Noyola, M. Dynamic equivalence between soft- and hard-core Brownian fluids. *Phys. Rev. E* **68,** 11405 (2003).

7. Heyes, D. M. & Brańka, A. C. The influence of potential softness on the transport coefficients of simple fluids. *J. Chem. Phys.* **122,** 234504 (2005).

8. Pond, M. J., Errington, J. R. & Truskett, T. M. Communication: Generalizing Rosenfeld's excess-entropy scaling to predict long-time diffusivity in dense fluids of Brownian particles: From hard to ultrasoft interactions. *J. Chem. Phys.* **134,** 81101 (2011).

9. Schmiedeberg, M., Haxton, T. K., Nagel, S. R. & Liu, A. J. Mapping the glassy

dynamics of soft spheres onto hard-sphere behavior. *EPL (Europhysics Lett.* **96,** 36010 (2011).

10. López-Flores, L., Ruíz-Estrada, H., Chávez-Páez, M. & Medina-Noyola, M. Dynamic equivalences in the hard-sphere dynamic universality class. *Phys. Rev. E - Stat. Nonlinear, Soft Matter Phys.* **88,** (2013).

11. Hansen, J. P. & McDonald, I. R. *Theory of Simple Liquids*. *StatMech* **2nd,** (2006).

12. Bernu, B., Hansen, J. P., Hiwatari, Y. & Pastore, G. Soft-sphere model for the glass transition in binary alloys: Pair structure and self-diffusion. *Phys. Rev. A* **36,** 4891–4903 (1987).

13. Roux, J. N., Barrat, J. L. & Hansen, J. P. Dynamical diagnostics for the glass transition in soft-sphere alloys. *J. Phys. Condens. Matter* **1,** 7171–7186 (1989).

14. Barrat, J. & Latz, A. Mode coupling theory for the glass transition in a simple binary mixture. *J. Phys. Condens. Matter* **2,** 4289 (1990).

15. Prestipino, S., Saija, F. & Giaquinta, P. V. Phase diagram of softly repulsive systems: The Gaussian and inverse-power-law potentials. *J. Chem. Phys.* **123,** (2005).

16. Barros De Oliveira, A., Netz, P. A., Colla, T. & Barbosa, M. C. Thermodynamic and dynamic anomalies for a three-dimensional isotropic core-softened potential. *J. Chem. Phys.* **124,** (2006).

17. Engel, M. & Trebin, H. R. Self-assembly of monatomic complex crystals and quasicrystals with a double-well interaction potential. *Phys. Rev. Lett.* **98,** (2007).

18. Krekelberg, W. P. *et al.* Generalized Rosenfeld scalings for tracer diffusivities in not-so-simple fluids: Mixtures and soft particles. *Phys. Rev. E - Stat. Nonlinear, Soft Matter Phys.* **80,** (2009).

19. Gallo, P. & Sciortino, F. Ising universality class for the liquid-liquid critical point of a one component fluid: A finite-size scaling test. *Phys. Rev. Lett.* **109,** (2012).

20. Nauroth, M. & Kob, W. Quantitative test of the mode-coupling theory of the ideal glass transition for a binary Lennard-Jones system. *Phys. Rev. E - Stat. Physics, Plasmas, Fluids, Relat. Interdiscip. Top.* **55,** 657–667 (1997).

21. Bennemann, C., Paul, W., Baschnagel, J. & Binder, K. Investigating the influence of different thermodynamic paths on the structural relaxation in a glass-forming polymer melt. *J. Phys. Condens. Matter* **11,** 2179–2192 (1999).

22. Pedersen, U. R., Bailey, N. P., Schrøder, T. B. & Dyre, J. C. Strong Pressure-Energy Correlations in van der Waals Liquids. *Phys. Rev. Lett.* **100,** 015701 (2008).

23. Bailey, N. P., Pedersen, U. R., Gnan, N., Schrøder, T. B. & Dyre, J. C. Pressure-energy correlations in liquids. I. Results from computer simulations. *J. Chem. Phys.* **129,** 184507 (2008).

24. Coslovich, D. & Roland, C. M. Thermodynamic Scaling of Diffusion in Supercooled Lennard-Jones Liquids. *J. Phys. Chem. B* **112,** 1329–1332 (2008).

25. Ingebrigtsen, T. S., Schrøder, T. B. & Dyre, J. C. What is a simple liquid? *Phys. Rev. X* **2,** 1–20 (2012).

26. Bacher, A. K., Schrøder, T. B. & Dyre, J. C. Explaining why simple liquids are quasi-universal. *Nat. Commun.* **5,** (2014).

27. Pedersen, U. R., Costigliola, L., Bailey, N. P., Schrøder, T. B. & Dyre, J. C. Thermodynamics of freezing and melting. *Nat. Commun.* **7,** (2016).

28. Schrøder, T. B., Pedersen, U. R., Bailey, N. P., Toxvaerd, S. & Dyre, J. C. Hidden scale invariance in molecular van der Waals liquids: A simulation study. *Phys. Rev. E* **80,** 041502 (2009).

29. Grzybowski, A. & Paluch, M. in *The Scaling of Relaxation Processes* (eds. Friederic, F. & Loidl, A.) (Springer International Publishing, 2018). doi:10.1007/978-3-319-72706-6

30. Tölle, A., Schober, H., Wuttke, J., Randl, O. G. & Fujara, F. Fast relaxation in a fragile liquid under pressure. *Phys. Rev. Lett.* **80,** 2374–2377 (1998).

31. Tölle, A. Neutron scattering studies of the model glass former ortho -terphenyl. *Reports Prog. Phys.* **64,** 1473 (2001).

32. Dreyfus, C. *et al.* Temperature and pressure study of Brillouin transverse modes in the organic glass-forming liquid orthoterphenyl. *Phys. Rev. E* **68,** 011204 (2003).

33. Casalini, R. & Roland, C. M. Thermodynamical scaling of the glass transition dynamics. *Phys. Rev. E - Stat. Physics, Plasmas, Fluids, Relat. Interdiscip. Top.* **69,** 3 (2004).

34. Dreyfus, C., Le Grand, A., Gapinski, J., Steffen, W. & Patkowski, A. Scaling the α-relaxation time of supercooled fragile organic liquids. *Eur. Phys. J. B* **42,** 309–319 (2004).

35. Alba-Simionesco, C., Cailliaux, A., Alegría, A. & Tarjus, G. Scaling out the density dependence of the α relaxation in glass-forming polymers. *Europhys. Lett.* **68,** 58–64 (2004).

36. Pawlus, S. *et al.* Temperature and volume effects on the change of dynamics in propylene carbonate. *Phys. Rev. E - Stat. Nonlinear, Soft Matter Phys.* **70,** (2004).

37. Casalini, R. & Roland, C. M. Scaling of the supercooled dynamics and its relation to the pressure dependences of the dynamic crossover and the fragility of glass formers. *Phys. Rev. B - Condens. Matter Mater. Phys.* **71,** (2005).

38. Reiser, A., Kasper, G. & Hunklinger, S. Pressure-induced isothermal glass transition of small organic molecules. *Phys. Rev. B - Condens. Matter Mater. Phys.* **72,** (2005).

39. Roland, C. M., Bair, S. & Casalini, R. Thermodynamic scaling of the viscosity of van der Waals, H-bonded, and ionic liquids. *J. Chem. Phys.* **125,** 124508 (2006).

40. Casalini, R. & Roland, C. M. An equation for the description of volume and

temperature dependences of the dynamics of supercooled liquids and polymer melts. *J. Non. Cryst. Solids* **353,** 3936–3939 (2007).

41. Bailey, N. P., Pedersen, U. R., Gnan, N., Schrøder, T. B. & Dyre, J. C. Pressure-energy correlations in liquids. II. Analysis and consequences. *J. Chem. Phys.* **129,** 184508 (2008).

42. Grzybowski, A., Koperwas, K. & Paluch, M. Scaling of volumetric data in model systems based on the Lennard-Jones potential. *Phys. Rev. E* **86,** 031501 (2012).

43. Ngai, K. L. & Paluch, M. Corroborative evidences of $TV^\gamma$ -scaling of the α-relaxation originating from the primitive relaxation/JG β relaxation. *J. Non. Cryst. Solids* **478,** 1–11 (2017).

44. Casalini, R. & Roland, C. M. Determination of the Thermodynamic Scaling Exponent for Relaxation in Liquids from Static Ambient-Pressure Quantities. *Phys. Rev. Lett.* **113,** 085701 (2014).

45. Bardic, V. Y. & Shakun, K. S. Investigations of the Steepness of a Repulsive Potential in Accordance with the Equation of State and Light-scattering Spectraitle. *Ukr. J. Phys.* **50,** 404 (2005).

46. Grzybowski, A., Paluch, M. & Grzybowska, K. Consequences of an equation of state in the thermodynamic scaling regime. *J. Phys. Chem. B* **113,** 7419–7422 (2009).

47. Grzybowski, A., Paluch, M., Grzybowska, K. & Haracz, S. Communication: Relationships between Intermolecular potential, thermodynamics, and dynamic scaling in viscous systems. *J. Chem. Phys.* **133,** (2010).

48. Grzybowski, A., Haracz, S., Paluch, M. & Grzybowska, K. Density Scaling of Supercooled Simple Liquids Near the Glass Transition. *J. Phys. Chem. B* **114,** 11544–11551 (2010).

49. Grzybowski, A., Grzybowska, K., Paluch, M., Swiety, A. & Koperwas, K. Density

scaling in viscous systems near the glass transition. *Phys. Rev. E* **83,** 041505 (2011).

50. Koperwas, K. *et al.* Glass-Forming Tendency of Molecular Liquids and the Strength of the Intermolecular Attractions. *Sci. Rep.* **6,** 36934 (2016).

51. Jorgensen, W. L., Maxwell, D. S. & Tirado-Rives, J. Development and Testing of the OPLS All-Atom Force Field on Conformational Energetics and Properties of Organic Liquids. *J. Am. Chem. Soc.* **118,** 11225–11236 (1996).

52. *GROMACS Reference Manual, Version 5.1, The GROMACS development teams at the Royal Institute of Technology and Uppsala University, Sweden, ftp://ftp.gromacs.org/pub/manual/manual-5.1-beta1.pdf.*